\documentclass[twocolumn]{aastex701}
\usepackage{pkgs}

\begin{document}
\title{Demagnifying gravitational lenses as probes of\\ dark matter structures and nonminimal couplings to gravity}
\author{Hong-Yi Zhang}
\affiliation{Tsung-Dao Lee Institute \& School of Physics and Astronomy, Shanghai Jiao Tong University, Shanghai 201210, China}
\email[show]{hongyi18@sjtu.edu.cn}

\begin{abstract}
Magnification of total image fluxes is typically considered a defining feature of gravitational microlensing. In contrast, I will show that nonminimal couplings to gravity can generate regions of negative gravitational potential curvature, giving rise to the distinctive possibility of demagnification. Such events, appearing as flux troughs in microlensing light curves, provide a direct probe of dark matter structures and, crucially, offer a means to disentangle nonminimal couplings to gravity from other astrophysical and cosmological models.
\end{abstract}

\section{Introduction}
Gravitational lensing refers to the deflection of light trajectories by massive objects. It has become one of the most powerful probes of astrophysical structures. For instance, it has been used to determine the mass distribution of galaxies and clusters, study large-scale structure, and detect exoplanets \citep{meneghetti2021introduction, Narayan:1996ba}. In this work, I propose to use microlensing---where one measures flux variations rather than resolving multiple images---to probe dark matter (DM) structures and possible nonminimal couplings (NMCs) to gravity.

The nature of DM remains one of the central open problems in modern physics. Well-motivated candidates include weakly interacting massive particles, axions, dark photons, primordial black holes, massive astrophysical compact halo objects, etc. (see \citep{Cirelli:2024ssz} for a comprehensive review). Microlensing has been applied to constrain the fraction of macroscopic DM structures relative to the total DM abundance \citep{Fairbairn:2017dmf, Croon:2020ouk, Croon:2020wpr, Fujikura:2021omw, Li:2025mqx, Schiappacasse:2021zlr, Yin:2024xov}.

Microlensing signals, and thus constraints on DM, can be affected by self-interactions or NMCs, since they modify the internal structure of DM objects. The effect is most significant when the Einstein radius, which determines the typical separation of lensed images, is comparable to the lens size. For example, quartic self-interactions of axions alter the profile and stability of axion stars, making microlensing constraints degenerate in the axion self-coupling, axion mass, and axion-star mass \citep{Fujikura:2021omw}. Recently, simulations suggest that NMCs (e.g., via dimension-4 operators) can generate regions of effective negative gravitational curvature in dense DM halos \citep{Chen:2024pyr}. This novel effect motivates the use of microlensing as a probe of NMCs.

For nonminimally gravitating DM, the Newtonian potential $\Phi$ sourced by a rest-mass distribution $\rho(\bx)$ obeys a modified Poisson's equation \citep{Bettoni:2011fs, Bettoni:2013zma, Zhang:2023fhs, Zhang:2024bjo}
\begin{align}
\label{poisson_eq}
\nabla^2 \Phi = 4 \pi G (\rho + \epsilon L^2 \nabla^2\rho) \equiv 4\pi G \rho_L ~,
\end{align}
where $L$ is a length scale characterizing the NMC strength, and $\epsilon = \pm 1$ encodes the sign. This modification arises from the low-energy effective description of dimension-4 operators coupling curvature tensors to DM fluids \citep{Bettoni:2011fs} or fields \citep{Zhang:2024bjo}. A brief review of its derivation is provided in appendix \ref{app:derivation}. Several astrophysical and cosmological observations can be used to constrain $L$. In particular, strong and weak lensing of galaxy clusters require $L \lesssim 100 \mathrm{kpc}$ \citep{Zamani:2024qbx}. The success of the cold DM paradigm on large-scale matter perturbations implies a stronger bound, $L \lesssim 60 \mathrm{pc}$ \citep{Zhang:2023fhs}. For vector DM coupled via $R_{\mu\nu} X^\mu X^\nu$, multimessenger observations of GW170817 and the gravitational Cherenkov effect yield an even tighter constraint, $L \lesssim 18  \mathrm{pc}$ \citep{Zhang:2023fhs}.

NMCs to gravity are generic in cosmology and particle physics. They arise naturally from quantum corrections and are required for the renormalization of field theories in curved spacetime \citep{Birrell:1982ix, Weinberg:1995mt, Callan:1970ze, Freedman:1974gs, Freedman:1974ze}. They have been studied in a wide range of contexts, including DM phenomenology \citep{Ivanov:2019iec, Barman:2021qds, Zhang:2023fhs, Davoudiasl:2024grq, Ozsoy:2023gnl, Sankharva:2021spi, Lebedev:2022vwf, Cata:2016dsg, Capanelli:2024pzd, Zhang:2023ktk}, inflation \citep{Faraoni:2000wk, Turner:1987bw, Golovnev:2008cf, Hertzberg:2010dc, He:2018gyf}, modified gravity \citep{Berti:2015itd, Clifton:2011jh}, electromagnetic fields \citep{Johnson:2023skw, Jana:2021lqe}, and dark energy \citep{Pan:2025psn, Ye:2024ywg, Wolf:2025jed}. Despite their ubiquity, the phenomenological effects of NMCs are often degenerate with those of other models or parameters. For instance, while NMCs can reproduce the apparent phantom crossing of dark energy hinted by baryon acoustic oscillation data \citep{DESI:2024mwx, DESI:2025zgx, Pan:2025psn, Ye:2024ywg, Wolf:2025jed}, the same data can also be explained by evolving dark energy or/and DM scenarios that do not involve NMCs \citep{Luu:2025fgw, Chen:2025wwn, Khoury:2025txd, Braglia:2025gdo}. Similarly, DM with NMCs to gravity can fit galactic rotation curves, the radial acceleration relation, halo core column densities, and galaxy-cluster pressure profiles, providing a competitive alternative to models invoking modified DM profiles or baryonic feedback \citep{Gandolfi:2021jai, Gandolfi:2022puw, Gandolfi:2023hwx}. Therefore, identifying observables or methods that can uniquely signal NMC effects is essential for breaking these degeneracies.

In this work, I will show that the modified Poisson’s equation \eqref{poisson_eq} implies the possibility of demagnification in microlensing events. Observation of such signals would constitute a distinctive signature of NMCs in microlensing light curves. In what follows, I will first briefly review the basics of gravitational lensing in the minimal weak-field gravity and then demonstrate how NMCs to gravity would lead to the demagnification of total fluxes.

\section{Microlensing in geometrical optics}
\begin{figure}
\centering
\includegraphics[width=\linewidth]{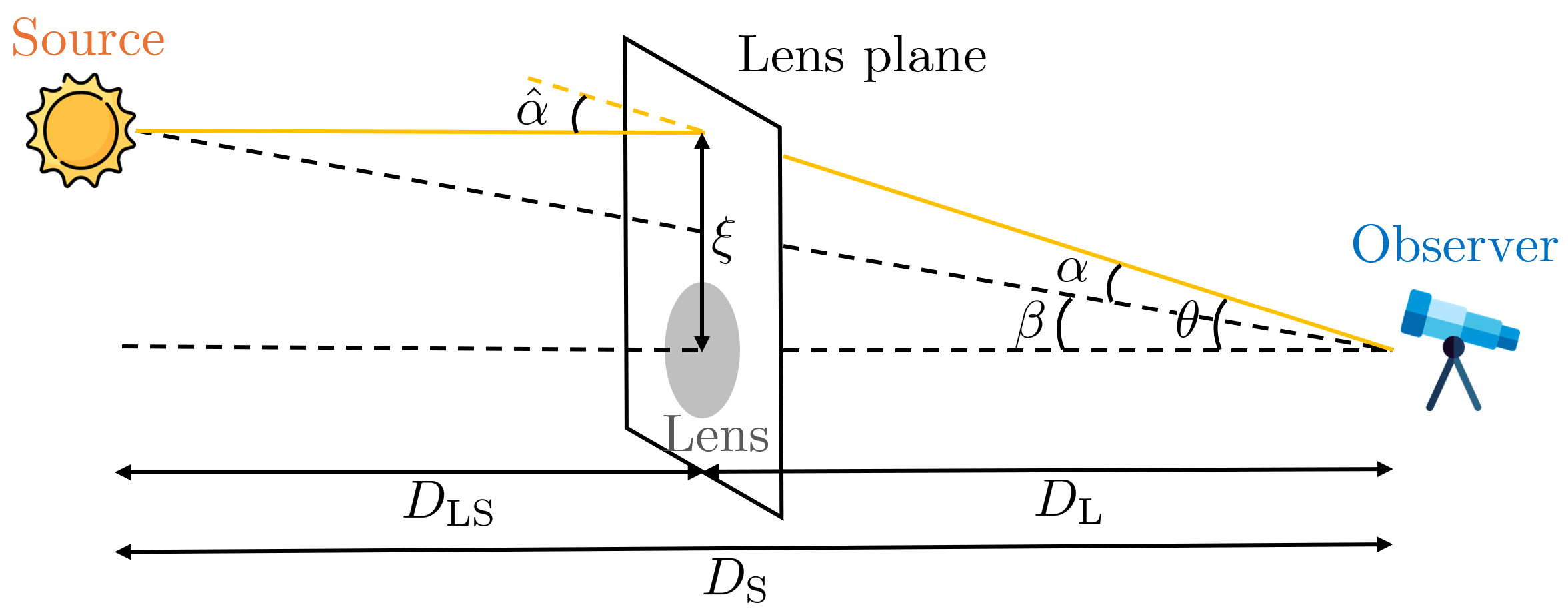}
\caption{A typical gravitational lensing system.}
\label{fig:lensing}
\end{figure}

A typical gravitational lensing system is schematically shown in figure \ref{fig:lensing}. Assuming a spherically symmetric lens with density distribution $\rho(r)$, the lens equation that relates different angles is
\begin{align}
\label{lens_eq}
\beta = \theta - \frac{D_\mrm{LS}}{D_\mrm{S}} \hat\alpha ~,
\end{align}
where $\beta$ and $\theta$ are the intrinsic and apparent angular position of the source relative to the optical axis connecting the lens and the observer, $D_\mrm{LS}$ is the distance between the lens and the source, $D_\mrm{S}$ is the distance between the source and the observer, and the deflection angle $\hat\alpha$, in the thin screen approximation, is given by \citep{meneghetti2021introduction, Narayan:1996ba}
\begin{align}
\label{lens_eq1}
\hat\alpha &= \frac{4 G M(\xi)}{\xi} ~,\\
\label{lens_eq2}
M(\xi) &= 4\pi \int_0^\xi \xi' d\xi' \int_0^\infty dz'~ \rho(r') ~,
\end{align}
with $\xi=D_\mrm{L} \theta$ being the impact parameter and $r'=({\xi'}^2 + {z'}^2)^{1/2}$. For a point lens, the solution for $\beta(\theta_\mrm{E}) = 0$ gives the Einstein angle
\begin{align}
\theta_\mrm{E} &= \sqrt{\frac{4GM D_\mrm{LS}}{D_\mrm{S} D_\mrm{L}}} \nonumber\\
&\sim 9\times 10^{-4} \arcsec \l( \frac{M}{M_\odot} \r)^{1/2} \l( \frac{10\kpc}{D_\mrm{S}} \r)^{1/2} ~,
\end{align}
where in the second equality I have assumed $D_\mrm{L} \sim D_\mrm{LS}$. Thus the angular separation between images due to lensing of stellar objects is on the order of milliarcseconds, far below the typical resolution of telescopes. Specifically, the ideal angular resolution of a telescope is $\theta \simeq \lambda/D \simeq 0.1 \arcsec (\lambda/520\mrm{nm}) (1\m/D)$, where $\lambda$ is the observed wavelength and $D$ is the telescope's aperture diameter. With the Einstein angle, it is useful to rewrite the lens equation in terms of dimensionless variables as
\begin{align}
\label{lens_eq3}
u = t - \frac{G(t)}{t} \sep
G(t) = \frac{M(\xi)}{M} ~,
\end{align}
where $u\equiv \beta/\theta_\mrm{E}$, $t\equiv \theta/\theta_\mrm{E}$, and $G(t)$ describes the distribution of the lens mass projected onto the lens plane, normalized by the total lens mass. The projected mass profile $G(t)$ represents the fraction of the lens mass enclosed within the impact parameter $\xi$. For a given source with angular location $u$, there could exist multiple images with different angular locations $t_i$.

Microlensing events are typically characterized by a transient amplification in the observed flux of lensed images. Since gravitational lensing conserves surface brightness, any change in the total flux is directly proportional to the ratio of the solid angles subtended by the image and the source. The magnification factor is therefore given by \citep{Croon:2020wpr}
\begin{align}
\label{magnification}
\mu &= \sum_i \abs{ \frac{\theta_i}{\beta} \frac{d\theta_i}{d\beta} } \nonumber\\
&= \sum_i \abs{ \l( 1 - \frac{G(t_i)}{t_i^2} \r) \l( 1 + \frac{G(t_i) - t_i G'(t_i)}{t_i^2} \r) }^{-1}  ~,
\end{align}
where the subscript $i$ denotes each solution of the lens equation. As the source moves relative to the line joining the observer and the lens, both the angular position $u$ and the magnification factor $\mu$ vary with time $\tau$. The impact parameter can be parameterized as
\begin{align}
u(\tau) = \sqrt{u_0^2 + \frac{\tau^2}{\tau_\mrm{E}}} ~,
\end{align}
where $u_0$ is the minimum angular separation between the source and the lens (attained at $\tau=0$), and $\tau_\mrm{E}$ is the Einstein time scale, defined by $\tau_\mrm{E} = D_\mrm{L}\theta_\mrm{E}/v$ with $v$ being the transverse velocity. This formulation enables the calculation of light curves that describe the temporal evolution of the magnification during a microlensing event. In microlensing surveys such as EROS-2 \citep{EROS-2:2006ryy}, Subaru HSC \citep{Niikura:2017zjd}, and MACHO \citep{MACHO:2000qbb}, a transit is counted as an event if the maximum magnification exceeds the threshold value $\mu\geq 1.34$.

As shown in equation \eqref{magnification}, the microlensing light curves are influenced by the lens's mass distribution. When the Einstein radius $R_\mrm{E} = D_\mrm{L} \theta_\mrm{E}$ is significantly larger than the physical size of the lens, the lens is effectively a point and any finite-size effects would be imperceptible. Conversely, when the lens size $R$ greatly exceeds $R_\mrm{E}$, the magnification is significantly suppressed. Therefore, to explore new physics effects in microlensing, it is essential to focus on gravitational lenses whose sizes are comparable to their Einstein radii, i.e., $R\sim R_\mrm{E}$ \citep{Croon:2020wpr, Croon:2020ouk, Fujikura:2021omw}.

\section{Demagnifying lenses}
The lensing formalism described in the previous section can be generalized to include NMC effects by replacing $\rho(r)$ with the effective density $\rho_L(r)$, which determines the Newtonian potential according to \eqref{poisson_eq}. To quantify the efficiency of demagnifying microlensing events, it is useful to define a threshold impact parameter $u_\mrm{T}$, below which a detectable reduction in magnification occurs. For instance, one may impose the condition 
\begin{align}
\label{uT}
\mu (u\leq u_\mrm{T}) \leq 0.9 ~,
\end{align}
corresponding to a $10\%$ reduction in total flux. For a pointlike lens, demagnification is impossible by construction, implying $u_\mrm{T}=0$. For an extended lens, $u_\mrm{T}$ depends on its mass density profile, as can be seen from \eqref{lens_eq3}. Demagnifying effects with nonzero $u_\mrm{T}$ require regions where the effective source $\rho_L(r)$ in the modified Poisson's equation becomes negative.\footnote{The regime $\rho_L<0$ corresponds to where the NMC correction becomes significant and may lead to demagnification signals. This does not imply to a direct violation of energy conditions in General Relativity, which are conditions on the energy-momentum tensor.}

For a spherically symmetric profile $\rho_L(r)$ up to a cutoff radius $r_\mrm{cut}$, characterized by a scale radius $r_\mrm{s}$ and density $\rho_\mrm{s}$, the projected mass profile in the dimensionless lens equation \eqref{lens_eq3} becomes
\begin{align}
G(t) = b \int_0^w w' dw' \int_0^{\sqrt{c^2 - {w'}^2}} dv' ~ \frac{\rho_L(s')}{\rho_\mrm{s}} ~,
\end{align}
where $b = 4 \pi r_\mrm{s}^3 \rho_\mrm{s}/M$ is an $\mcal O(1)$ coefficient, $w = \xi/r_\mrm{s}=t R_\mrm{E}/r_\mrm{s}$ is the dimensionless impact parameter, $c = r_\mrm{cut}/r_\mrm{s}$ is the concentration parameter, $s' =({w'}^2 + {v'}^2)^{1/2}$, and the effective mass density $\rho_L$ in terms of the dimensionless coordinate $s'$ is given by
\begin{align}
\rho_L(s') = \rho(s') + \epsilon \frac{L^2}{r_\mrm{s}^2} \nabla_{s'}^2 \rho(s') ~.
\end{align}
Given a rest-mass density profile $\rho(r)$, one can then solve the lens equation \eqref{lens_eq3} and calculate the magnification factor using \eqref{magnification}.

\begin{figure}
\centering
\includegraphics[width=\linewidth]{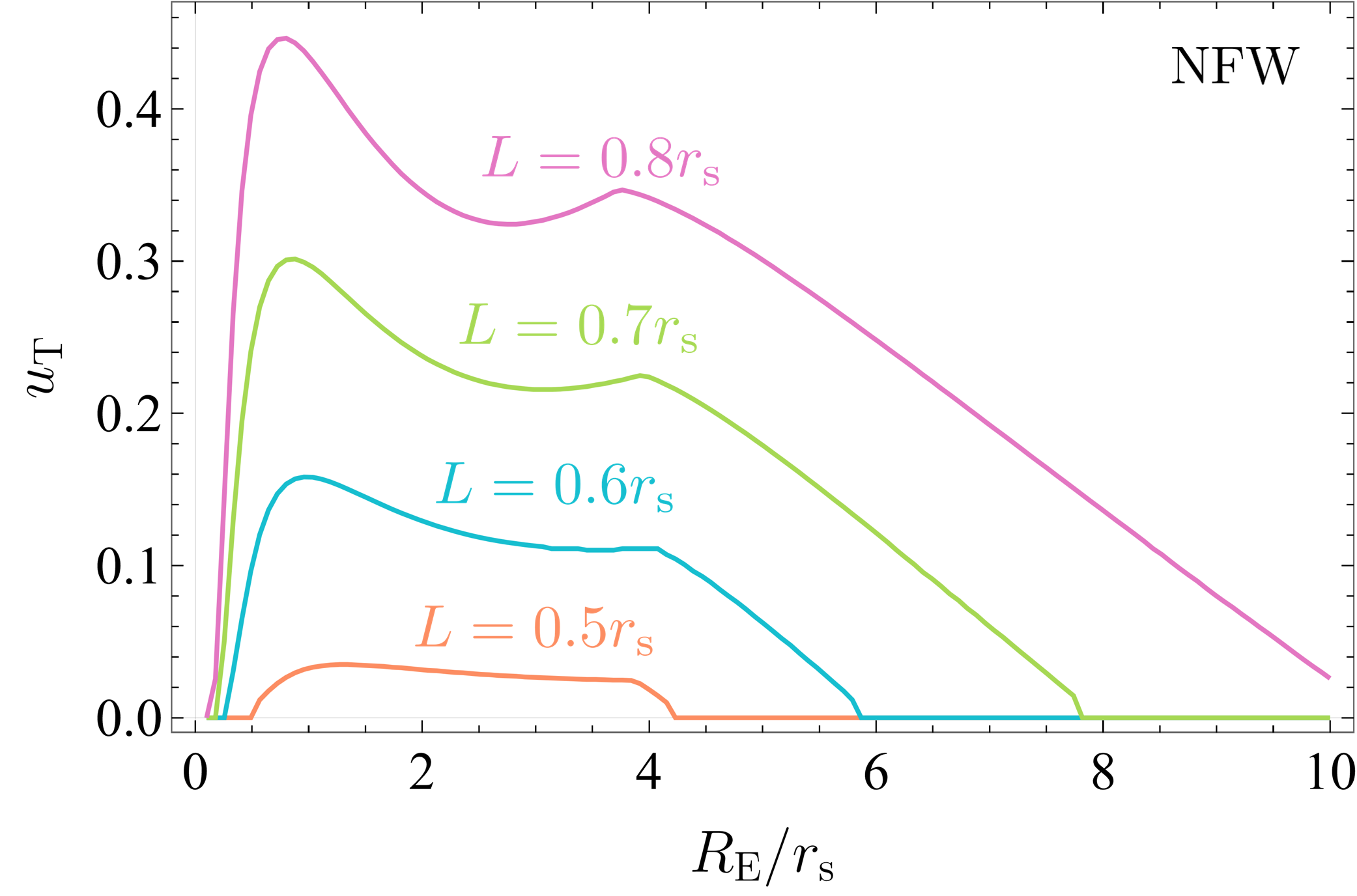}
\caption{Threshold impact parameter $u_\mrm{T}$ for the NFW profile, defined in equation \eqref{uT}, for different NMC strengths. It represents the angular separation between the source and the lens below which the total magnification is reduced by at least $10\%$. Here the NMC is taken with sign $\epsilon=-1$, allowing regions of negative effective density $\rho_L$ to form.}
\label{fig:utnfw}
\end{figure}

As a concrete example, let us consider gravitational lenses with the Navarro–Frenk–White (NFW) profile \citep{Navarro:1995iw}
\begin{align}
\label{profile_nfw}
\rho(r) = \frac{\rho_\mrm{s}}{(r/r_\mrm{s}) (1+r/r_\mrm{s})^2} ~.
\end{align}
This profile is commonly used to describe the spherically averaged distribution of virialized DM halos. As the mass integral of the NFW profile diverges logarithmically at large radii, it is customary to introduce a cutoff at the virial radius. Motivated by numerical simulations of axion miniclusters \citep{Eggemeier:2019khm}, here I adopt a truncation at $c=100$, and thus the prefactor in front of the integral is approximately $b \simeq 0.3$ for $L\leq r_\mrm{s}$. Following this, the threshold impact parameter $u_\mrm{T}$ for the NFW profile can be calculated and is shown in figure \ref{fig:utnfw}. As expected, $u_\mrm{T}\rightarrow 0$ in the limit $R_\mrm{E}/r_\mrm{s} \gg 1$, where the lens approaches the point-mass regime. For $R_\mrm{E}/r_\mrm{s} \ll 1$, the finite size of the lens suppresses gravitational lensing effects, rendering (de)magnification negligible. In the intermediate regime $R_\mrm{E}/r_\mrm{s}\sim 1$, finite-size effects become pronounced, and demagnification could occur. We see that the threshold impact parameter $u_\mrm{T}$ increases with the strength of the NMC, and $u_\mrm{T}\rightarrow 0$ if $L\ll r_\mrm{s}$. Therefore, DM structures of smaller sizes can be used to probe smaller NMC length $L$.

\begin{figure}
\centering
\includegraphics[width=\linewidth]{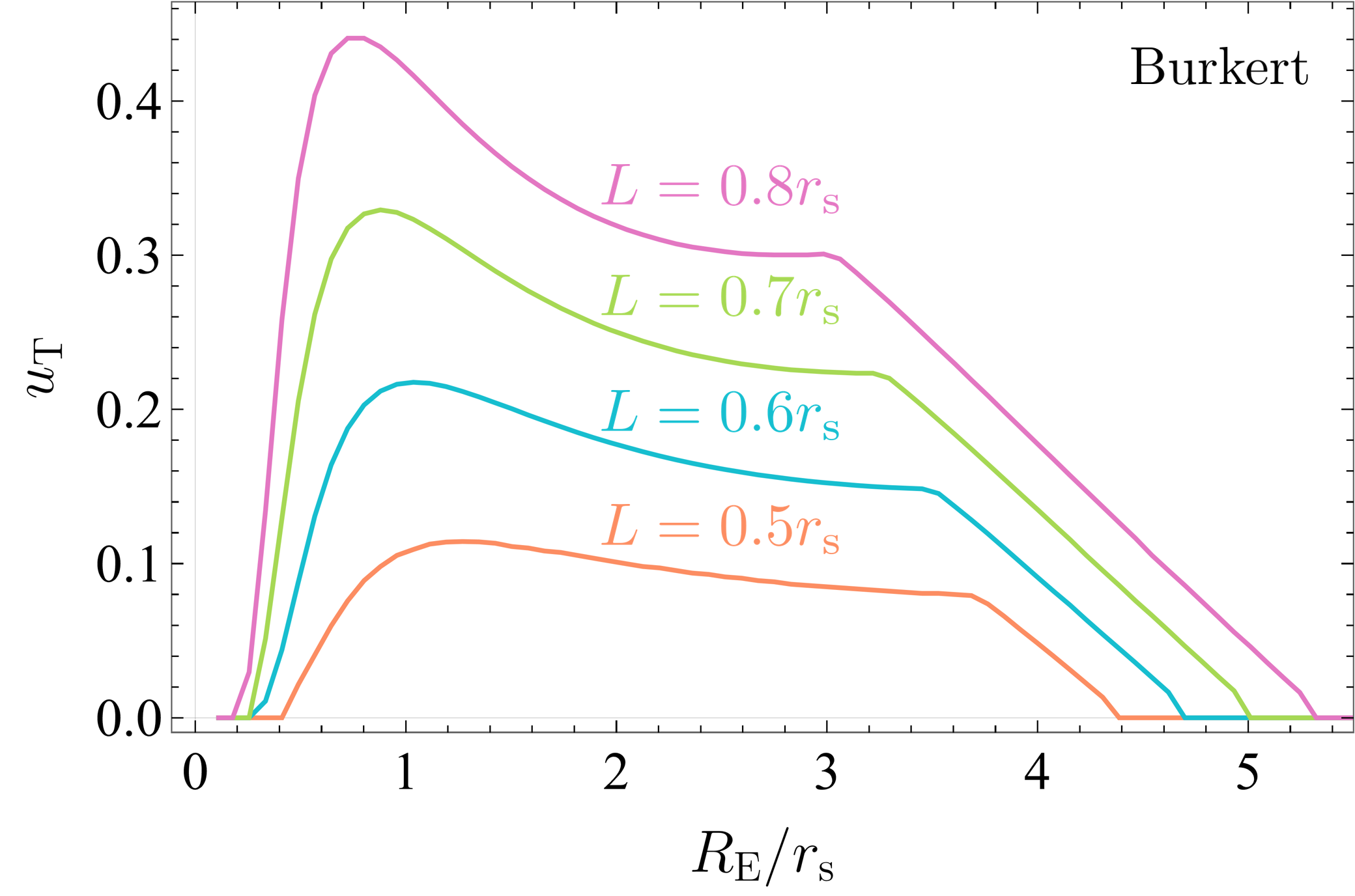}
\caption{Threshold impact parameter $u_\mrm{T}$ for the Burkert profile for different NMC strengths, using the same notation as in figure \ref{fig:utnfw}.}
\label{fig:utburkert}
\end{figure}

An alternative DM profile with qualitatively different central behavior is the Burkert profile, which features a flat core \citep{Burkert:1995yz},
\begin{align}
\rho(r) &= \frac{\rho_\mrm{s}}{(1 + r/r_\mrm{s}) [1 + (r/r_\mrm{s})^2]} ~.
\end{align}
This profile is motivated by observed galactic rotation curves that indicate the presence of cores \citep{Li:2020iib} and by numerical simulations showing that baryonic feedback can reduce central densities \citep{Mollitor:2014ara}. As with the NFW profile, a cutoff radius at $c = 100$ is imposed to avoid divergence at large radii and to facilitate comparison with the NFW results. For this profile, one finds $b \simeq 0.3$ for $L \lesssim 50 r_\mrm{s}$. The corresponding threshold impact parameter $u_\mrm{T}$ is illustrated in figure \ref{fig:utburkert}.

\begin{figure*}
\centering
\includegraphics[height=0.246\linewidth]{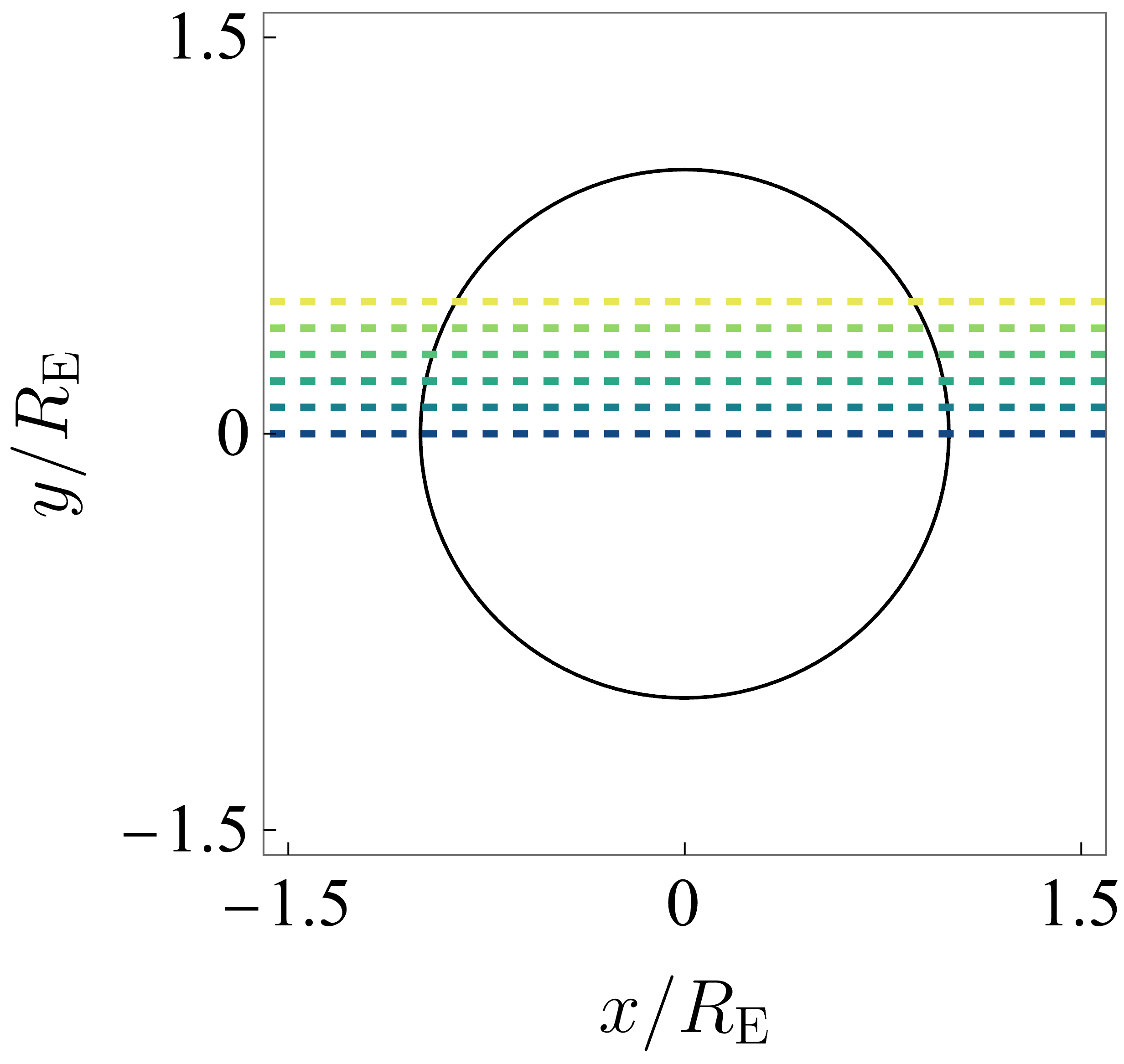}
\includegraphics[height=0.247\linewidth]{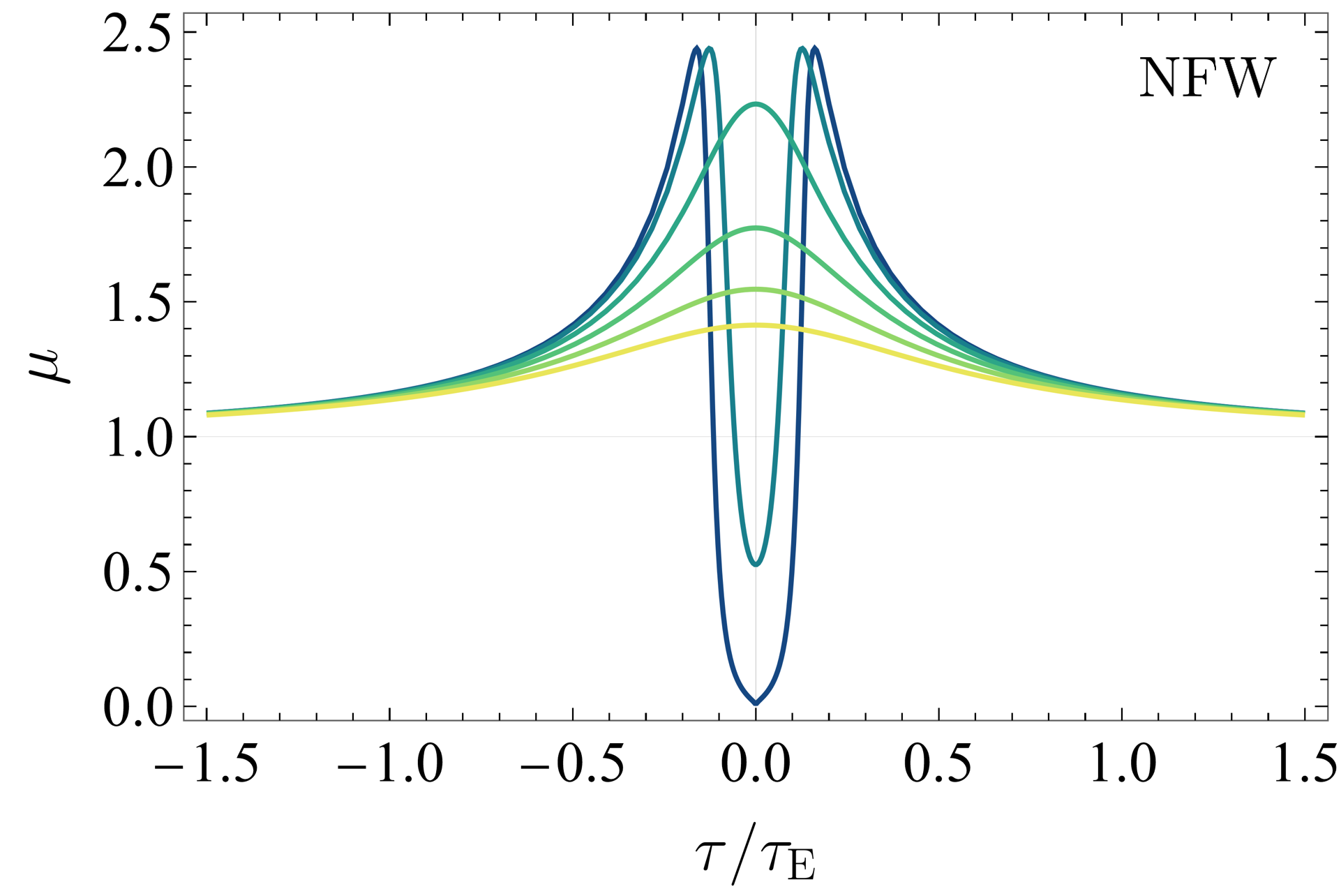}
\includegraphics[height=0.247\linewidth]{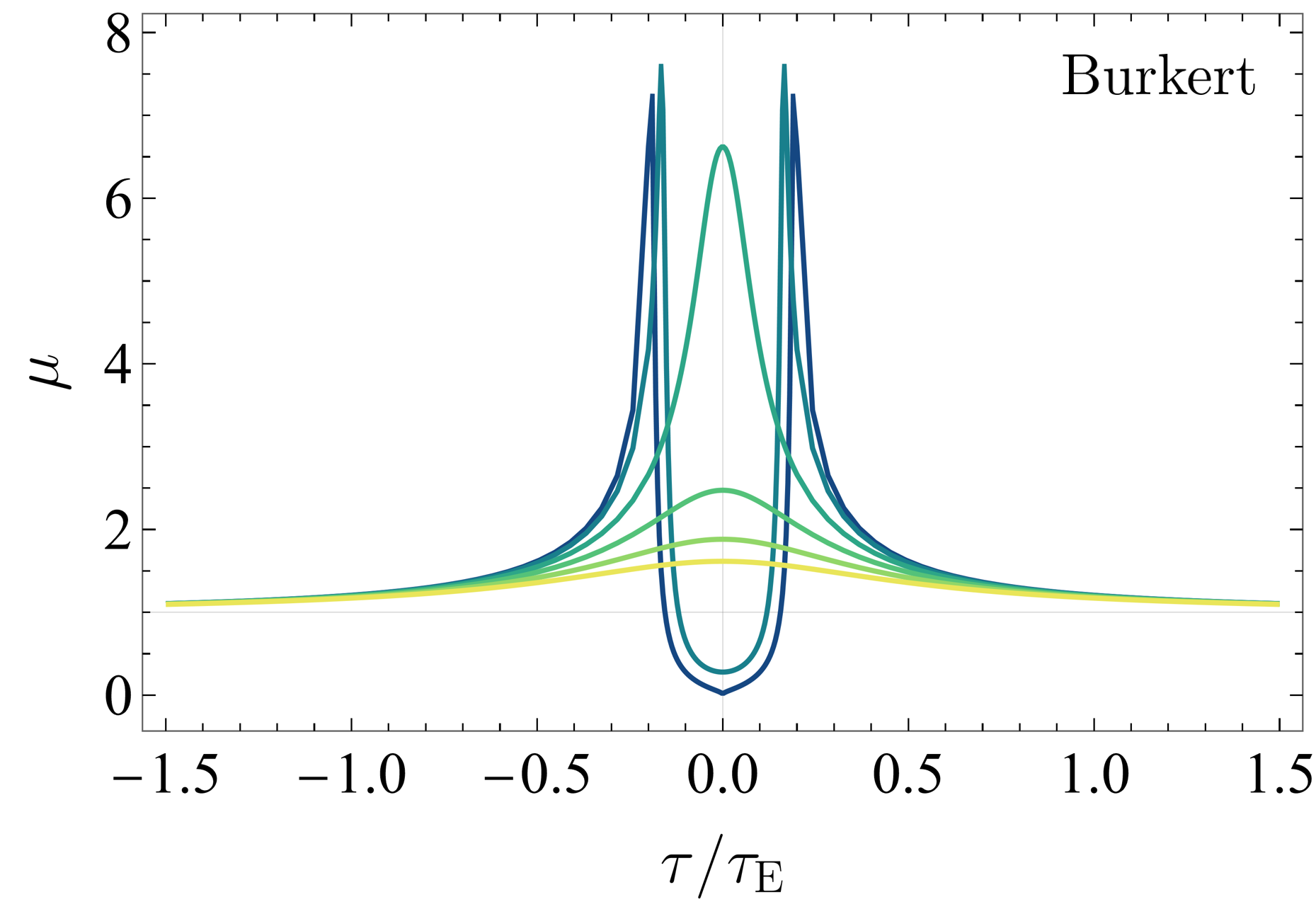}
\caption{Light curves for microlensing by NFW (middle) and Burkert (right) lenses, with source trajectories shown on the lens plane (left). Dashed lines in the left panel indicate trajectories with different minimum angular separations $u_0$, from 0 (blue) to 0.5 (yellow), while the black circle marks the Einstein ring. In the middle and right panels, magnification troughs appear near $\tau = 0$ for small $u_0$ in both the NFW and Burkert profiles, illustrating the demagnifying effect caused by NMCs. Here, parameters are set to $\epsilon = -1$, $L = 0.6 r_\mathrm{s}$, and $R_\mathrm{E}/r_\mathrm{s} = 3$.}
\label{fig:lightcurve}
\end{figure*}

In figure \ref{fig:lightcurve}, light curves for both the NFW and Burkert profiles are presented. In the left panel, dashed lines indicate different source trajectories on the lens plane, corresponding to various values of the minimum angular separation $u_0$, while the black circle represents the Einstein ring. The middle and right panels show the evolution of the magnification for the NFW and Burkert profiles, respectively, with NMC strength set to $L = 0.6 r_\mathrm{s}$ and the ratio of Einstein radius to scale radius $R_\mathrm{E}/r_\mathrm{s} = 3$. As seen in the middle and right panels, significant flux reduction ($>10\%$) occurs near $\tau = 0$ when $u_0 < 0.2$, consistent with figures \ref{fig:utnfw} and \ref{fig:utburkert}. Away from $\tau = 0$, two magnification peaks appear, so a demagnifying event is typically characterized by a central trough flanked by two peaks.

The final example considered is extended macroscopic DM objects with the density profile \citep{Fujikura:2021omw}
\begin{align}
\label{profile_soliton}
\rho(r) = \rho_\mrm{s} \l( 1 + \frac{r}{r_\mrm{s}} \r)^2 e^{-2r/r_\mrm{s}} ~,
\end{align}
where $\rho_\mrm{s} = M/(7\pi r_\mrm{s}^3)$ and $M$ is the total mass. This profile could describe a class of DM solitons that are generically predicted both analytically and in numerical simulations for bosonic dark matter \citep{Zhou:2024mea, Liebling:2012fv, Visinelli:2021uve, Lee:1991ax}, including axion stars \citep{Visinelli:2017ooc}, boson stars \citep{Visinelli:2021uve}, oscillons \citep{Zhang:2020bec, Zhang:2020ntm, Zhang:2021xxa}, and Q-balls \citep{Zhou:2024mea}. In such cases, the soliton radius and mass are related by $r_\mrm{s} \simeq 47.1 \MP^2/ (m^2 M)$, where $m$ is the constituent particle mass \citep{Zhang:2024bjo}.\footnote{To convert the relation reported in \citep{Zhang:2024bjo} to this form, I used $r_\mrm{s}\simeq 0.239 r_{95}$ according to \eqref{profile_soliton}, where $r_\mrm{95}$ is the $95\%$-mass radius.} For DM solitons, a minimum radius exists, given by $L / r_{\mathrm{s,min}} \approx 0.6$ independent of the NMC sign $\epsilon$ \citep{Zhang:2024bjo}. Calculating the threshold impact parameter $u_\mathrm{T}$ shows that even solitons with radii approaching $r_{\mathrm{s,min}}$ do not produce demagnification. However, solitons are expected to reside at the centers of DM halos or minihalos \citep{Schive:2014dra, Chen:2020cef, Amin:2022pzv, Levkov:2018kau}, whose overall density is well described by the Burkert profile \citep{Schive:2014dra}. Such host halos can still generate demagnifying microlensing signals.

\section{Discussions}
NMCs to gravity are widely invoked in cosmology and particle physics, but their observational signatures are often degenerate with other parameters or models, making them difficult to isolate. In this work, I have shown that demagnifying signals in gravitational microlensing---manifesting as flux troughs in light curves in the geometrical optical regime---provide a distinctive probe of DM structures and NMCs to gravity. Similar effects may also appear in strong or weak lensing regimes, which warrant further study.

The possibility of demagnification may also arise from wave-optical effects or exotic matter. For monochromatic light, interference patterns can appear in gravitational lensing when the wavelength is comparable to twice the Schwarzschild radius of the lens \citep{Matsunaga:2006uc}. This condition enforces a specific source-lens correlation and limits its event rate. More importantly, in realistic situations the source typically emits a continuous spectrum, and the frequency dependence of the wave effect suppresses  demagnification in the light curves \citep{Bai:2018bej, Tamta:2024pow}. Demagnification can also occur for lenses composed of exotic matter, such as Ellis wormholes supported by a phantom scalar field \citep{Abe:2010ap, Tsukamoto:2017hva, Takahashi:2013jqa, Kitamura:2012zy}. Leaving aside the likelihood of detecting these signals, the resulting light curves in both cases remain qualitatively distinct from those produced by NMCs.

Observable demagnification effects requre an intermediate regime in which finite-lens effects become important, $R\sim R_\mrm{E}$, and the NMC scale is not much smaller than the lens size, $L\gtrsim R$. This has two important implications for observations and phenomenology: If a lens with $R \sim R_\mrm{E}$ is detected without any demagnification effects, this can be translated into a constraint on the NMC coupling scale $L$, assuming an NFW or Burkert profile; (ii) If a demagnifying signal is observed in lensing surveys, this would indicate the existence of a lens that could be overlooked in searches relying solely on magnifying templates. For example, the observation of an object with mass $\sim 10^6 M_\odot$ and radius $\sim 80 \mrm{pc}$ at a distance $\sim \mrm{Gpc}$ (with Einstein radius $\sim 10 \mrm{pc}$) in strong lensing \citep{Powell:2025rmj} could imply a constraint $L \lesssim 80 \mrm{pc}$ if it is confirmed as an NFW/Burkert DM halo in the future. The demagnification signals may be suppressed due to finite-source or blending effects, which should be minimized. Finite-source effects average the (de)magnification over the source surface and therefore smooth both peaks and troughs. Blending arises from unresolved background light and reduces the observed fractional deviation by a factor $F_\mrm{s}/(F_\mrm{s}+F_\mrm{b})$, where $F_\mrm{s,b}$ are fluxes of the source and the blended background.

Current and previous gravitational lensing surveys generally assume positive magnification, either adopting a magnification threshold of 1.34 (e.g., EROS-2 \citep{EROS-2:2006ryy}, Subaru HSC \citep{Niikura:2017zjd}, MACHO \citep{MACHO:2000qbb}) or applying light-curve templates derived from point lenses (e.g., OGLE-IV \citep{udalski2015ogleivfourthphaseoptical, Mroz:2017mvf}). Consequently, demagnifying lenses may have been overlooked in past analyses. With machine learning increasingly used for signal identification, methods have been developed to detect microlensing from both pointlike and extended lenses, with the latter typically producing three flux peaks \citep{CrispimRomao:2024nbr}. Extending these approaches to include light-curve templates featuring flux troughs could be a useful direction.

In summary, NMCs modify the effective source in the Poisson's equation for the gravitational potential, allowing regions of negative effective density that induce defocusing of light rays. This produces a qualitatively distinct microlensing signature: a flux deficit below the unlensed baseline for sufficiently small impact parameters. Observing such features could constrain NMC strengths or reveal objects that might be overlooked in previous surveys. Developing and implementing new templates in gravitational lensing surveys, therefore, would open a new avenue for testing the fundamental properties of DM and the gravitational interaction on subgalactic scales.

\begin{acknowledgments}
I would like to thank Minxi He and Tucker Manton for carefully reading the draft. HYZ acknowledges support from the Shanghai Magnolia Plan Pujiang Program 25PJA073, CPSF General Program 2025M783412, and NSFC Grant No. 12350610240.
\end{acknowledgments}

\appendix
\section{Derivation of the modified Poisson's equation}
\label{app:derivation}

The modified Poisson's equation \eqref{poisson_eq} can be derived by modeling dark matter either as a perfect fluid \citep{Bettoni:2011fs, Bettoni:2013zma} or as a field \citep{Zhang:2023fhs, Zhang:2024bjo}. These two descriptions are often used interchangeably, particularly for wavelike dark matter \citep{Hui:2021tkt}. In this section, I briefly review the derivation assuming a scalar-field description for simplicity. The same conclusion holds if dark matter consists of multiple field components or behaves as a fluid.

Let us start with a free scalar field $\f$ with NMCs of the lowest mass dimension
\begin{align}
\label{action_phi}
S = \int d^4x \sqrt{-g} \bigg[ \frac{\MP^2}{2} R - \frac{1}{2} \pd_\mu\f \pd^\mu\f - \frac{m^2}{2}\f^2 - \frac{\xi}{2} R\f^2 \bigg] ~,
\end{align}
where $\MP = (8\pi G)^{-1/2}$ is the reduced Planck mass, $m$ is the field mass, and $\xi$ parameterizes the NMC strength. NMC terms involving spacetime derivatives are suppressed since dark matter is nonrelativistic. In the nonrelativistic limit, the field oscillates predominantly with frequency $\simeq m$. It is thus useful to expand the field as
\begin{align}
\label{nr_expansion_phi}
\f(t,\bx) = \frac{1}{\sqrt{2m}} \bigg[ e^{-imt} \psi(t,\bx) + e^{imt} \psi^*(t,\bx) \bigg] ~,
\end{align}
where $\psi$ varies on timescales much longer than $m^{-1}$. Accordingly, its temporal and spatial derivatives are small compared to $m$, and we define the small parameters
\begin{align}
\epsilon_t \equiv \abs{ \frac{\pd_t\psi}{m\psi} } \sep
\epsilon_x \equiv \abs{ \frac{\nabla^2\psi}{2m^2\psi} } ~.
\end{align}
A slowly varying field such as $\psi$ can be further decomposed into modes with different frequencies,
\begin{align}
\label{mode_expansion_psi}
\psi = \sum_{\nu=-\infty}^{\infty} \psi_\nu e^{i\nu m t} ~,
\end{align}
where $\psi_\nu$ vary slowly in time and $\psi_0$ dominates over higher modes. For dark matter to remain nonrelativistic, gravitational interactions must be weak. Adopting the Newtonian gauge,
\begin{align}
\label{metric}
ds^2 = -(1+2\Phi) dt^2 + (1-2\Psi) \delta_{ij} dx^i dx^j ~,
\end{align}
we introduce three additional small parameters,
\begin{align}
\epsilon_\Phi \equiv \abs{2\Phi} \sep
\epsilon_\xi \equiv \abs{ \frac{\xi R}{2m^2} } \sep
\epsilon \equiv \abs{ \frac{\psi}{\sqrt{m} \MP} } ~.
\end{align}
To leading order in $\mcal O(\epsilon)$, where $\epsilon$ collectively denotes all small parameters, the two scalar potentials can be set equal, since $\Phi - \Psi \sim \epsilon_\psi^2$ \citep{Salehian:2021khb}. Substituting the expansions \eqref{nr_expansion_phi}, \eqref{mode_expansion_psi} and the metric \eqref{metric} into the action \eqref{action_phi}, one obtains a hierarchy of equations for the modes that can be solved order by order. At leading order in $\mcal O(\epsilon)$, the effective action reduces to
\begin{align}
S_\mrm{NR} = \int d^4x \bigg[& \MP^2 \Phi \nabla^2\Phi + i\dot\psi \psi^* + \frac{(\nabla^2\psi)}{2m} \psi^* \nonumber \\
&- m\Phi|\psi|^2 - \frac{\xi}{m}(\nabla^2\Phi) |\psi|^2 \bigg] ~.
\end{align}
This yields the modified Poisson equation \eqref{poisson_eq}, with $\xi/m^2 \rightarrow \epsilon L^2$ (here $\epsilon$ stands for the sign of the NMC). In addition, the dark matter field $\psi$ satisfies the Schroedinger equation, which can be recast into fluid continuity and Euler equations via the Madelung transformation \citep{Hui:2021tkt}.

\bibliography{ref}
\bibliographystyle{aasjournalv7}



\end{document}